\documentclass[aps,prl,twocolumn,superscriptaddress,preprintnumbers]{revtex4-2}
\pdfoutput=1
\usepackage{amsmath}
\usepackage{graphicx}
\usepackage{bbm}
\usepackage{amssymb}
\usepackage{enumitem}
\usepackage{booktabs}
\usepackage{verbatim} 
\usepackage{soul}
\usepackage{mathrsfs}
\usepackage[linktocpage=true,
colorlinks=true,
urlcolor=blue,
anchorcolor=blue,
citecolor=blue,
filecolor=blue,
linkcolor=blue,
menucolor=blue,
]{hyperref}

\newcommand{\ex}{\mathrm{e}}

\newcommand{\dd}{\mathrm{d}}
\newcommand{\R}{\mathbb{R}}

\newcommand{\hook}{\mathbin{\rule[.2ex]{.4em}{.03em}\rule[.2ex]{.03em}{.9ex}}}

\usepackage{color}

\def\nn{\nonumber}
\newcommand{\ii}{\mathrm{i}}

\newcommand{\Z}{\mathbb{Z}}

\newcommand{\cA}{\mathcal{A}}
\newcommand{\cF}{\mathcal{F}}
\newcommand{\cG}{\mathcal{G}}
\newcommand{\cI}{\mathcal{I}}
\newcommand{\cK}{\mathcal{K}}
\newcommand{\cN}{\mathcal{N}}
\newcommand{\cQ}{\mathcal Q}

\newcommand{\cD}{\mathcal{D}}

\newcommand{\chispinor}{\zeta}
\newcommand{\ca}{\sigma}

\newcommand{\mf}[1]{\mathfrak{#1}}

\newcommand{\bW}{\mathbb{W}}
\newcommand{\bT}{\mathbb{T}}
\newcommand{\ttw}{\texttt{w}}
\newcommand{\ttt}{\texttt{t}}

\begin{document}

\title{
The superconformal index and localizing higher derivative supergravity}

\author{Florian Gaar}
\affiliation{Mathematical Institute, University of Oxford, Woodstock Road, Oxford, OX2 6GG, U.K.}

\author{Jerome P. Gauntlett}
\affiliation{Abdus Salam Centre for Theoretical Physics, Imperial College, Prince Consort Road, London, SW7 2AZ, U.K.}

\author{Jaeha Park}
\affiliation{Abdus Salam Centre for Theoretical Physics, Imperial College, Prince Consort Road, London, SW7 2AZ, U.K.}

\author{James Sparks}
\affiliation{Mathematical Institute, University of Oxford, Woodstock Road, Oxford, OX2 6GG, U.K.}

\begin{abstract}
\noindent 
We show how equivariant localization can be used to compute the on-shell action for supersymmetric 
$D=5$ $AdS$ rotating, charged black holes in theories of supergravity with higher derivatives. An exact match
with a dual field theory computation of the superconformal index in a Cardy-like limit is achieved.

\end{abstract}

\maketitle

\enlargethispage{0.25\baselineskip}

\section{Introduction}\label{sec:intro}
The $AdS$/CFT correspondence provides key insights into the quantum properties of black holes and hence quantum gravity. 
A particularly interesting setting is supersymmetric $AdS_5$ solutions of string or M-theory that are
dual to $d=4$, $\mathcal{N}=1$ superconformal field theories (SCFTs).
In this context, on the gravity side it is expected that the full supersymmetric partition function,
with specific $AdS_5$ boundary conditions,
should be equal to the superconformal index of the dual SCFT.
In the large $N$ limit the superconformal index has various complex saddles, which includes
an important  saddle that can be used to compute the black hole entropy after taking a Legendre transform. 
This saddle can be isolated by taking a certain Cardy-like limit 
and then taking the large $N$ limit \cite{Choi:2018hmj,Kim:2019yrz,
Cabo-Bizet:2019osg,Cassani:2021fyv,ArabiArdehali:2021nsx}.

More precisely, consider a
4d $\cN=1$ SCFT on $S^1 \times S^3$,
whose symmetry algebra contains $J_1, J_2$
generating the Cartan sub-algebra of the $SO(4)$ symmetry of the round $S^3$,
and
conserved charges $Q_I$ ($I = 1, \dots, n+1$) made of linear combinations of the superconformal R-charge and $n$ Abelian flavour charges.
Choosing a complex supercharge $\cQ$ satisfying
$[ J_1 , \cQ ] = [J_2 , \cQ ] = \frac{1}{2} \cQ$, $[Q_I , \cQ] = - r_I \cQ$,
the superconformal index is defined as the supersymmetric partition function on $S^1 \times S^3$,
\begin{align}
	\cI (\omega_1, \omega_2, \varphi^I) = {\rm Tr}\mskip2mu \big[\ex^{-\beta \{ \cQ, \overline{\cQ} \} + \omega_1 J_1 + \omega_2 J_2 + \varphi^I Q_I }\big] \,,
\end{align}
where $\omega_1$, $\omega_2, \varphi^I$ are chemical potentials for the charges $J_1,J_2, Q_I$, respectively,
subject to the constraint
\begin{align}\label{multi_chempot_constraint_v1}
	\omega_1 + \omega_2 - 2 r_I \varphi^I = \pm 2 \pi \ii \,.
\end{align}
Extending \cite{Choi:2018hmj,Kim:2019yrz,
Cabo-Bizet:2019osg,Cassani:2021fyv,ArabiArdehali:2021nsx,Ohmori:2021dzb},
it has been shown in \cite{Cassani:2024tvk} that the index in the Cardy-like limit $\omega_{1},\omega_2\to 0$, has the form
\begin{align}\label{index}
	- \log \cI = \frac{k_{IJK} \varphi^I \varphi^J \varphi^K}{6\mskip1mu \omega_1 \omega_2} - k_I \varphi^I \frac{\omega_1^2 + \omega_2^2 - 4\pi^2}{24\mskip1mu \omega_1 \omega_2}  \,,
\end{align}
where $k_{IJK} = {\rm Tr}\mskip2mu (Q_I Q_J Q_K)$ and $k_I = {\rm Tr}\mskip2mu (Q_I)$ are cubic and linear 't Hooft anomaly coefficients, respectively.
In \eqref{index} we have neglected terms that are exponentially suppressed in the Cardy-like limit, as well as a possible 
$\log |\cG|$ term \footnote{Here $|\cG|$ is the order of a discrete one-form symmetry group, if such a symmetry exists, which would lead to a multiplicity of the saddle.}.

For holographic theories, the cubic anomaly term in~\eqref{index} dominates at large $N$. Using two-derivative $D=5$
supergravity, it has been shown in \cite{Cabo-Bizet:2018ehj,Cassani:2019mms} that the on-shell action for a complex locus of supersymmetric but non-extremal black holes,  
precisely agrees with the leading term in \eqref{index}. 
Moreover, by taking a Legendre transform 
one obtains \cite{Hosseini:2017mds} the Bekenstein--Hawking entropy of the supersymmetric black holes of \cite{Gutowski:2004yv,Cvetic:2004ny}.

Recently there has been significant further progress in matching the sub-leading terms in \eqref{index} to a gravitational
computation. Since we are considering sub-leading terms in $N$ one necessarily needs to carry out the computation
in the context of a $D=5$ supergravity theory with higher derivatives. The result \eqref{index} depends on 't Hooft anomaly coefficients
and these are holographically determined from the Chern--Simons terms \cite{Witten:1998qj}
\begin{equation}\label{CSterms}
	\frac{1}{24\pi^2}(k_{IJK}\cA^I\wedge \cF^J\wedge \cF^K-\frac{1}{8}k_I\cA^I\wedge \mathcal{R}_{ab}\wedge \mathcal{R}^{ab})\, ,
\end{equation}
where $\cA^I$ are Abelian gauge fields dual to the R-symmetry and flavour symmetry currents with $\cF^I=\dd\cA^I$, and 
$\mathcal{R}_{ab}$ is the Riemann curvature two-form. One thus needs a $D=5$ supergravity with four derivatives that
provides a supersymmetrization of \eqref{CSterms}. 
The approach of \cite{Bobev:2021qxx,Bobev:2022bjm,Cassani:2022lrk,Cassani:2024tvk} is to 
utilise conformal supergravity, which provides a powerful
framework to deal with higher derivative theories 
\footnote{$\alpha'$ corrections to the entropy for specific $AdS_5\times S^5$ black holes were shown to vanish directly in $D=10$ \cite{Melo:2020amq}.}. 
A key ingredient is the identification of a supersymmetric invariant, involving
the supersymmetric completion of the square of the Weyl tensor, that contains 
the mixed gauge-gravitational Chern--Simons term \cite{Hanaki:2006pj}. 
Various off-shell invariants contribute to the chiral anomaly; the proposal in \cite{Cassani:2024tvk} is to
modify the two-derivative theory and combine it with an action built from the Weyl invariant.

This approach was used for minimal supergravity in \cite{Bobev:2022bjm,Cassani:2022lrk}
to match the universal part of \eqref{index}, with flavour chemical potentials turned off. 
With vector multiplets, a matching was made in \cite{Cassani:2024tvk},
focussing on equal angular momentum, certain theories and also resorting to numerics.
Given the topological nature and 
simplicity of the expression \eqref{index}, 
one might expect \cite{Cassani:2024tvk}
that it can be derived in a more robust and universal way using the techniques of equivariant localization in supergravity \cite{BenettiGenolini:2023kxp}. Indeed this has been achieved for the leading order term in 
\cite{Colombo:2025ihp,Colombo:2025yqy,BenettiGenolini:2025icr}.

The purpose of this paper is to show that localization can also be used to obtain the sub-leading terms.
We utilise two ingredients: first, that localization
of $D=5$ supergravity at the two-derivative level can be carried out by performing a dimensional reduction and then utilising
results for $D=4$, $\mathcal{N}=2$ supergravity \cite{BenettiGenolini:2025icr,BenettiGenolini:2024xeo, BenettiGenolini:2024lbj}. Second, that localization can be extended to
off-shell conformal supergravity in $D=4$ and hence to higher derivative theories~\cite{BenettiGenolini:2026qdm}. The strategy is then clear: we consider the
four-derivative supergravity  theory in $D=5$ of \cite{Cassani:2024tvk}, dimensionally reduce to $D=4$, and then use the results of \cite{BenettiGenolini:2026qdm}.
We show that \eqref{index} can be recovered without requiring an explicit solution, 
without having to assume equal angular momenta,  and
for a general class of theories.
Our procedure just assumes that a solution with certain prescribed global data exists
\footnote{We also assume that any such solution, which in general is complex, is a \emph{bona fide} saddle point of the Euclidean gravitational path integral for computing the superconformal index; for some further
discussion see \cite{BenettiGenolini:2026raa,Krishna:2026rma}.}.

\section{Higher derivative supergravity}
The $D=5$ theory of \cite{Cassani:2024tvk} is constructed using off-shell conformal supergravity, with suitable compensator multiplets and gauge-fixing 
also required in order to recover Poincar\'e supergravity. The $D=5$ superconformal multiplets consist of the Weyl multiplet coupled to
$n$ vector multiplets, one of which is a compensator, with an additional linear multiplet compensator, as in \cite{Ozkan:2013nwa}.
The two-derivative Lagrangian $\mathcal{L}_{2\partial}$ is controlled by a totally symmetric tensor $C_{IJK}$ associated with the very special geometry of the scalar manifold. 
The effective
off-shell
action considered
in \cite{Cassani:2024tvk} has the form 
\begin{align}
	\label{eq:effective 4deriv lagrangian}
	\mathcal{L} = \mathcal{L}_{2\partial}\big|_{C_{IJK}\rightarrow C_{IJK}+\alpha \tilde{\lambda}_{IJK}} + \alpha\mskip1mu \mathcal{L}_{C^2}(\lambda_I)\,,
\end{align}
where $\mathcal{L}_{C^2}$ is the Weyl-squared Lagrangian 
\cite{Hanaki:2006pj} involving the vector multiplets, with $\lambda_I$ a coupling,
 and one works to first order in the parameter $\alpha$.
 The coupling $\tilde{\lambda}_{IJK}$ is a symmetric tensor which, 
in combination with $C_{IJK}$ and $\lambda_I$, 
 ensures that the theory can 
 reproduce the anomalies of the dual field theory \cite{Cassani:2024tvk}. After gauge fixing we obtain 
 parameters $\zeta_I$ that determine the gauging. To simplify subsequent formulae we set the gauge coupling  
 $\mf{g}=1$.

By considering the resulting Chern--Simons terms that appear in the gauge-fixed action
 one can identify 
\cite{Cassani:2024tvk} the cubic and linear 't Hooft anomaly coefficients  
of the 4d SCFT with gravitational variables via
 \begin{equation}\label{AdS/CFT}
	k_{IJK} = \frac{\pi}{4G_{(5)}}C_{IJK}^{(\alpha)} \,,\qquad k_I = - \frac{24\pi}{G_{(5)}} \alpha \lambda_I \,,
\end{equation}
 where
 \begin{equation}\label{CS_corrected}
	C_{IJK}^{(\alpha)} \equiv C_{IJK} + \alpha \tilde{\lambda}_{IJK} - 8 \alpha  \lambda_{(I} \zeta_J \zeta_{K)} \,,
\end{equation}
and $G_{(5)}$ is the $D=5$ Newton constant \footnote{One can gauge fix in $D=5$ and then multiply by an overall factor of 
$\frac{1}{16\pi G_{(5)}}$.}.
Notice in particular that $k_I$ is determined purely by the Weyl-squared Lagrangian. By contrast, 
$k_{IJK}$ gets contributions from $\mathcal{L}_{2\partial}$, the Weyl-squared Lagrangian and from other $D=5$ invariants (e.g. see table 1 of \cite{Hristov:2025ygn}) which are effectively incorporated together in $ \tilde{\lambda}_{IJK} $. 
We also note that we have rescaled $C_{IJK}$ in \cite{Cassani:2024tvk} via $C_{IJK}\to \frac{1}{6}C_{IJK}$ to agree with the notation of
\cite{BenettiGenolini:2025icr}; in particular, working to first order in $\alpha$, the scalars 
$Y^I$ in the $D=5$ vector multiplets are constrained via $\frac{1}{6}C_{IJK}Y^I Y^J Y^K=1$.
Finally, as we will see, we can identify $ \frac{1}{2} \zeta_I$ with $r_I$ appearing in the $\cN=1$ algebra, $[Q_I, \cQ] = - r_I \cQ$.

Supersymmetric configurations of the $D=5$ theory have a supersymmetric conformal Killing vector $\cK$ that can be constructed 
as a bilinear in the Killing spinors. After gauge fixing this becomes a Killing vector.
We are interested in solutions in which there is another Killing vector, $\ell$,
which is used to carry out a  Kaluza--Klein
(KK) dimensional reduction to $D=4$, following \cite{BenettiGenolini:2025icr}.
We introduce  a local coordinate so that we can write $\ell = \partial_{x^5}$ and introduce the 
KK ansatz for the metric 
$\dd s^2_{(5)} = \ex^{-4\lambda}(\dd x^5 - A^0)^2 + \ex^{2\lambda} \dd s^2_{(4)}$. 
We can also write the $D=5$ scalars and gauge fields in the form $Y^I=-\ex^{2\lambda}z_2^I$ 
(so $\ex^{6\lambda}$ is determined by the cubic constraint for $Y^I$),
$\mathcal{A}^I=A^I+z^I_1(\dd x^5 - A^0)+a^I \dd x^5$, where $A^I$ are $D=4$ gauge fields with $F^I=\dd A^I$.
The $a^I$ are constants which are defined up to some constant shift of the variables;
quantities that are invariant are defined with a caron, e.g. $\check z_1^I=z_1^I+a^I$, $\check A^I=A^I+a^I A^0$, $\check F^I=F^I+a^I F^0$.
The resulting $D=4$ theory then has $n+2$ vector multiplets, one of which can be viewed as a compensator multiplet, 
with $n+2$ gauge fields $A^\Lambda\equiv (A^0, A^I)$ and $n+1$ complex scalars $z^I \equiv z_1^I+\ii z_2^I$. The latter are associated with the
constrained vector multiplet scalars $X^\Lambda=(X^0,X^I)$, 
where $z^I=X^I/X^0$.

By carrying out the dimensional reduction as discussed in \cite{Butter:2014iwa}, from \eqref{eq:effective 4deriv lagrangian}
we obtain a $D=4$ off-shell $\cN=2$
conformal supergravity theory that is constructed from a Weyl multiplet, $n+1$ vector multiplets and
a T-log multiplet. 
The resulting $D=4$ theory can be expressed in terms of a prepotential $\cF$, given by
\begin{align}	\label{modified prep multicharge}
	\cF(X^\Lambda,A_{\bW^2},A_\bT) = &\  (C_{IJK} + \alpha  \tilde{\lambda}_{IJK}) \frac{\check{X}^I \check{X}^J \check{X}^K}{6X^0}\nn \\
	&+ \alpha \lambda_I \frac{\check{X}^I}{X^0} ({ \ttw A_{\bW^2} + \ttt A_\bT}) \,,
\end{align}
where $X^\Lambda,A_{\bW^2}$ and $A_\bT$ are the lowest component fields of the 
vector, Weyl-squared and T-log multiplets, respectively.
The T-log multiplet is constructed from the vector multiplet containing $X^0$ \cite{Butter:2014iwa}, which is a constrained combination of a chiral and anti-chiral multiplet of weight $w'=1$.
Observe that $\cF$ is homogeneous
of degree 2 with respect to $X^\Lambda$, and degree 1 with respect to $A_{\bW^2}$ and $A_\bT$.
In our normalizations  (as in \cite{BenettiGenolini:2026qdm}), 
we have $\ttw=\frac{1}{32}$ and $\ttt=\frac{1}{3}$, but it is illuminating to 
keep $\ttw$ and $\ttt$ 
to see how  
various terms contribute to the localization 
computation.  

\section{Localization}
In \cite{BenettiGenolini:2024xeo, BenettiGenolini:2024lbj} 
it was shown that the on-shell action for Euclidean supersymmetric solutions to $D=4$, 
$\mathcal{N}=2$ gauged supergravity coupled to vector multiplets, at the level of two derivatives,  
localizes to the fixed points of the supersymmetric 
Killing vector $\xi$.  Here $\xi$ is constructed 
as a bilinear in the Killing spinors. Furthermore, all boundary terms in the action
precisely cancel (after an appropriate Legendre transform ensuring supersymmetry), with the result expressed only in terms of the prepotential $\mathcal{F}$ evaluated at the fixed points, 
together with the weights of $\xi$  and 
certain other topological data. 
Some key localization results of \cite{BenettiGenolini:2024xeo, BenettiGenolini:2024lbj} have recently been extended  
to conformal supergravity constructions of general higher derivative 
theories in $D=4$, including for prepotentials of the form~\eqref{modified prep multicharge} 
\cite{BenettiGenolini:2026qdm}\footnote{For the higher derivative theories, the boundary terms are under less control.}.

Thus, we can use the localization formulas of \cite{BenettiGenolini:2024xeo, BenettiGenolini:2024lbj, BenettiGenolini:2026qdm} to evaluate 
the $D=5$ on-shell action via the KK reduction described at the end of the last section. 
There are several subtleties.
First, here we only consider solutions where we can assume the $D=5$ gauge fields 
are globally defined one-forms, which avoids issues 
with gauge-dependence and an additional Chern--Simons-like 
contribution to the action. 
Second, we may circumvent the thorny issue 
of boundary terms and holographic renormalization in $D=5$ by using a background subtraction prescription
implemented in $D=4$, as discussed in \cite{BenettiGenolini:2025icr}. 
Third, the localization computation in \cite{BenettiGenolini:2026qdm} is for a $D=4$ Euclidean theory with independent
chiral and anti-chiral fields which were taken to be real. For our purposes, we need to consider a suitable
analytic continuation which in practice means considering prepotentials $\mathcal{F}^\pm$ which are functions
of the chiral and anti-chiral fields, which are no longer real, with 
$\cF^\pm(X^\Lambda_\pm,A_{\bW^2_\pm},A_{\bT_\pm}) \equiv \pm \ii \cF(X^\Lambda_\pm,A_{\bW^2_\pm},A_{\bT_\pm})$
\footnote{Additional discussion of reality conditions in localization appear in \cite{BenettiGenolini:2024lbj,BenettiGenolini:2025icr,BenettiGenolini:2026qdm}.}.

The $D=4$ supersymmetric conformal Killing vector $\xi$ 
is simply the projection of the $D=5$ vector $\mathcal{K}$ 
\cite{BenettiGenolini:2025icr}. We denote the $D=4$ 
spacetime by $M_4$. 
We will only need to consider the case 
where $\xi$ has isolated fixed points on $M_4$, which are called ``nuts'' \cite{Gibbons:1979xm}. These are labelled nut$_\pm$, 
where the $D=4$ Killing spinor necessarily has either positive/negative chirality $\chi=\pm 1$ at each nut. 
The tangent space at such a point is $\R^4/\Z_d=(\R^2\oplus\R^2)/\Z_d$, 
allowing for orbifold points of order $d\in\mathbb{N}$, 
and the weights of $\xi$ on each of the $\R^2_i/\Z_d$ factors are denoted 
$b_i$, $i=1,2,$ respectively.
 The arguments of $\mathcal{F}$ in \eqref{modified prep multicharge}
are rescaled by certain spinor bilinears $S_\pm$ in the localization formulas of \cite{BenettiGenolini:2026qdm}, 
correlated with anti-chiral/chiral multiplets, respectively,
 in such a way that 
these also have simple localization properties at the fixed points.
In particular, 
only $A_{\bW^2_\mp}, A_{\bT_\mp}$ contribute to the fixed point formula 
at a nut$_\pm$, respectively, via the key formulas \cite{BenettiGenolini:2026qdm} 
\begin{align}\label{AWAT}
S^2_\pm A_{\mathbb{W}^2_\mp}|_{\text{nut}_\pm}& =16(b_1\pm b_2)^2\, ,\nonumber\\
S^2_\pm A_{\mathbb{T}_\mp}|_{\text{nut}_\pm}& =\frac{w'}{2}(b_1\mp b_2)^2\, ,
\end{align}
and recall 
from above 
that $w'=1$.
Likewise, the following 
``dressed'' versions of the scalars 
$X^\Lambda$ (or their gauge-invariant versions $\check{X}^I\equiv X^I+a^I X^0$)  appear
in the fixed point formula 
\begin{align}\label{PhiLambda}
\Phi^\Lambda_0|_{\text{nut}_\pm} = \mp 2S_\pm X^\Lambda_\mp ,\,
\end{align}
and we similarly define
$\check\Phi^I_0 \equiv \Phi^I_0 +a^I \Phi^0_0$. 
The $\Phi^\Lambda_0$ in \eqref{PhiLambda} are precisely the zero-form components of  equivariantly closed forms associated to the gauge field curvatures:
\begin{align}\label{Fluxdxi}
\dd_\xi (F^\Lambda + \Phi_0^\Lambda) = 0\, ,
\end{align}
with $\dd_\xi = \dd - \xi \hook$ the equivariant exterior derivative, and we follow the normalizations in \cite{BenettiGenolini:2026qdm}.

With all these definitions in hand, we can finally state the  fixed point formula \cite{BenettiGenolini:2026qdm}   for the $D=5$ on-shell action associated to the prepotential \eqref{modified prep multicharge}: 
\begin{align}\label{I4FP_multi}
	&I^{\rm FP}_{4\partial}  = \frac{\ii \pi^2}{2 G_{(5)}}\sum_{\rm nuts_\pm} \frac{1}{d} \Bigg\{ (C_{IJK} + \alpha \tilde{\lambda}_{IJK})\frac{ \check{\Phi}_0^I  \check{\Phi}_0^J\check{\Phi}_0^K}{6b_1 b_2 \Phi_0^0} \nn \\
	& + 4 \alpha  \lambda_I \frac{  \check{\Phi}_0^I }{b_1 b_2 \Phi_0^0}\Big[16\ttw (b_1\pm b_2)^2  + \frac{1}{2}\ttt (b_1\mp b_2)^2\Big] \Bigg\}\,,
\end{align}
where we have used \eqref{AWAT}. 
Here ${\Phi}^\Lambda_0, b_1,b_2$ and $d$ depend on the nut; when we need to make this clear, 
we label the nuts by an index $a$ and write 
${\Phi}_0^\Lambda|_a, b_1^a,b_2^a,d^a$. 
The $\check{\Phi}^I_0$ are then subject to the constraint
\begin{equation}\label{Phiconstraint}
-2Q^{(\ell)}\Phi^0_0|_a+ \zeta_I \check\Phi_0^I |_a = \sigma_a (b_1^a - \chi_a b_2^a) \,,
\end{equation}
at each fixed point \cite{BenettiGenolini:2025icr}. Here  $\chi_a=\pm 1$ is the chirality of the $a$th nut$_\pm$, 
$\sigma_a$ is another sign that we discuss momentarily, 
and $Q^{(\ell)}$ is the charge of the $D=5$ Killing spinor 
under the KK vector $\ell$.  

Finally, we will also use the following localization formula
for the flux, via \eqref{Fluxdxi}:
\begin{align}\label{fluxlocalize}
\frac{1}{4\pi}\int_{D}F^\Lambda = \frac{1}{2}\left(\frac{1}{d_Nb_N}\Phi^\Lambda_0|_N + \frac{1}{d_Sb_S}\Phi^\Lambda_0|_S\right)\, ,
\end{align}
where $D\subset M_4$ is a two-dimensional subspace to which 
$\xi$ is tangent and non-zero. Necessarily then 
$D\cong \mathbb{WCP}^1_{[d_N,d_S]}$ is a weighted projective 
space/spindle, and the weights are $b_N=-b_0/d_N$, $b_S=b_0/d_S$ where $\xi |_D= b_0\partial_{\phi}$ with 
$\phi$ having period $2\pi$. 

\section{Black holes}
We now evaluate the on-shell action for a class of Euclidean black holes, 
using the formula \eqref{I4FP_multi}. We regularize the action using 
background subtraction, subtracting the on-shell action of Euclidean $AdS_5$, as described in \cite{BenettiGenolini:2025icr}. 
As advertised in the introduction, we do this without 
using any explicit black hole solutions; all we need is certain global data, which we now summarize, together with the assumption that such  supergravity solutions exist.

The Euclidean black holes have topology 
$M_5=\R^2\times S^3$, with conformal boundary $\partial M_5=S^1_\tau\times S^3$ 
where the superconformal index is defined. 
We introduce polar coordinates 
$(r,\tau)$ on the $\R^2$ factor, where $\tau$ 
parametrizes the thermal circle and has period $2\pi$, while the horizon 
$S^3$ is at $r=0$. The generators of the 
Cartan subgroup $U(1)^2\subset SO(4)$ acting on $S^3$ are denoted
 $\partial_{\phi_i}$, $i=1,2$, so that $(\tau,\phi_1,\phi_2)$ 
parametrize the torus $U(1)^3$. In this basis the $D=5$
supersymmetric conformal Killing vector is
\begin{align}\label{KSUSY}
\mathcal{K} = \partial_\tau +\varepsilon_1\partial_{\phi_1}+
\varepsilon_2\partial_{\phi_2} = (1,\varepsilon_1,\varepsilon_2)\, ,
\end{align}
where \cite{BenettiGenolini:2025icr} defined $\mathcal{K}$ as a bilinear in the 
$D=5$ Killing spinors. In \eqref{KSUSY}, and subsequently, we identify 
vector fields with elements of the Lie algebra $\mathtt{u}(1)^3\cong \R^3$. 

A special role is played by codimension two subspaces $\cD_a\subset M_5$ that are fixed by $U(1)\subset U(1)^3$ subgroups. Writing the generating vector fields as $V_a\in\Z^3$, we have:
\begin{align}
\cD_0 & \cong \R^2\times S^1_1\, ,  \quad  V_0  \equiv (0,0,1)\, ,\nonumber\\
\cD_1 & \cong S^3\, ,  \quad \qquad \ \mskip2mu  V_1 \equiv (1,0,0)\, ,\nonumber\\
\cD_2 & \cong \R^2\times S^1_2\, , \quad V_2  \equiv (0,1,0)\, ,
\end{align}
where $S^1_i$ is parametrized by $\phi_i$.
These describe the ``rod structure'' of \cite{Cassani:2025iix}, where note that $\cD_1$ is the horizon.
Importantly, $M_5$ has no non-trivial two-cycles, 
implying that the Abelian gauge fields $\mathcal{A}^I$ 
may be represented by global one-forms. In this global gauge, the $D=5$ Killing spinor $\chispinor$, which is charged 
under the combination $\zeta_I\mathcal{A}^I$, satisfies
\begin{align}\label{KScharge}
\mathcal{L}_{V_a}\chispinor = \frac{\ii}{2}\ca_a \chispinor\, .
\end{align}
Here $\ca_a=\pm 1$ for each $a=0,1,2$ simply follows from regularity 
of the spinor along each subspace $\cD_a$.

The conformal boundary is $\partial M_5=S^1_\tau\times S^3$,  
at constant radius $r\rightarrow\infty$. Euclidean global $AdS_5$ has 
the same conformal boundary, but different topology  $N_5=S^1_\tau\times \R^4$, 
where $\R^4$ fills the $S^3$. Using this for background subtraction 
effectively leads \cite{BenettiGenolini:2025icr} to the compact glued manifold 
$\overline{M}_5 \equiv M_5 \cup_{S^1_\tau\times S^3}(-N_5)\cong S^5$, which 
is topologically a five-sphere
\footnote{The gluing construction can be generalized to $\partial M_5 = S^1_\tau \times M_3$, where $M_3$ are squashed three-spheres/lens spaces \cite{Park:2025fon}.}.

\section{Reduction to $D=4$}
To reduce the black hole to $D=4$  we need to pick a Kaluza--Klein (KK)  circle. 
The final $D=5$ on-shell action must be independent of this choice, 
and this is a non-trivial consistency check of our computation.
We take the KK circle action to be an arbitrary $U(1)_\ell\subset U(1)^3$, 
specified by the primitive vector
\begin{align}\label{ell}
\ell = (-p,n_1,n_2)\in\Z^3\, ,
\end{align} 
where $\mathrm{gcd}(p,n_1,n_2)=1=\mathrm{gcd}(n_1,n_2)$ \footnote{The latter condition may be dropped, at the expense of introducing 
	factors of $\mathrm{gcd}(n_1,n_2)$ below.}. 
From~\eqref{KScharge} we have $\mathcal{L}_\ell \zeta = 
{\ii}Q^{(\ell)}\zeta$, and hence we read off the charge
\begin{align}
Q^{(\ell)} = \frac{1}{2}(-p \ca_1+n_1 \ca_2 + n_2 \ca_0)\, .
\end{align}

The $D=4$ spacetime is $M_4 =M_5/U(1)_\ell$, with 
projection  $\pi:M_5\rightarrow M_4$. 
This inherits a $U(1)^2\cong U(1)^3/U(1)_\ell$ action 
from $M_5$, which makes it toric. 
The $D=4$ supersymmetric Killing vector is the projection 
of \eqref{KSUSY}
\begin{align}
\xi\equiv \pi_*\mathcal{K}\, .
\end{align}
To apply the $D=4$ localization formula
\eqref{I4FP_multi}
 we need to compute the 
weights $(b^a_1,b^a_2)$ of $\xi$ at the fixed points $x_a\in M_4$. 
In $M_5$, each such fixed point defines a 
 circle fibre  $S^1_a=\pi^{-1}(x_a)$, and 
we also define the weights $w_a$ via
\begin{align}
\mathcal{K}|_{S^1_a} = -w_a \ell |_{S^1_a}\, .
\end{align}
This determines the zero-form part of the equivariantly closed form associated to the KK gauge field  \cite{BenettiGenolini:2025icr}
\begin{align}\label{Phi00}
\Phi^0_0|_{a} = w_a\, .
\end{align}

To compute all this data, we note that  
$M_4$ is topologically the total space of the complex line orbibundle
$\mathcal{O}(-p)\rightarrow \mathbb{WCP}^1_{[n_1,n_2]}$ over 
a spindle. 
The glued manifold $S^5/U(1)_\ell=\overline{M}_5/U(1)_\ell \cong 
\mathbb{WCP}^2_{[p,n_1,n_2]}$ is a weighted projective space. 
In terms of weighted homogeneous coordinates $(z_*,z_0,z_1)$ on this weighted projective space, the fixed 
points of $U(1)^2$ are $x_a=\{z_b=0 \, |\,  b\neq a\}$, for $a=0,1,*$, 
where $x_*$ corresponds to the origin of Euclidean global $AdS_5$. 
These are orbifold points of order $d^a$, with $d^0=n_1$, $d^1=n_2$, 
$d^*=p$. 
The toric geometry of this example was studied in \cite{BenettiGenolini:2024hyd}, and we may simply write down the weights
\begin{align}\label{weights}
(b^0_1,b^0_2,w_0) & = \frac{1}{n_1}(n_2\varepsilon_1-n_1\varepsilon_2, -n_1-p\mskip1mu\varepsilon_1,-\varepsilon_1)\, ,\nonumber\\
(b^1_1,b^1_2,w_1) &= \frac{1}{n_2}(-n_2-p\mskip1mu\varepsilon_2,n_1\varepsilon_2-n_2\varepsilon_1,-\varepsilon_2)\, ,\nonumber\\
(b^*_1,b^*_2,w_*) &= \mskip4mu\frac{1}{p}\mskip2mu(-n_2-p\mskip1mu\varepsilon_2,-n_1-p\mskip1mu\varepsilon_1,\mskip1mu 1 \mskip1mu)\, .
\end{align}

Next we turn to the fluxes. There is a single compact two-cycle in
 $M_4$, which is the projection of the black hole horizon $D_1\equiv\cD_1/U(1)_\ell \cong 
\mathbb{WCP}^1_{[n_1,n_2]}$. 
We define the gauge field fluxes through this cycle as
\begin{align}\label{flux}
\mathfrak{p}^\Lambda & \equiv \frac{1}{4\pi}\int_{{D}_1} F^\Lambda \nonumber\\
& = \frac{1}{2(n_2\varepsilon_1-n_1\varepsilon_2)}\left( \Phi^\Lambda_0|_0-\Phi^\Lambda_0|_{1}\right)\, ,
\end{align}
where in the second equality we used the localization formula \eqref{fluxlocalize}, together with the relevant weights in 
\eqref{weights}. Notice that using \eqref{Phi00}, \eqref{weights} the flux of the KK 
gauge field is $\mathfrak{p}^0=-1/(2n_1n_2)$, which is the correct Chern number 
for the fibration $S^3\rightarrow \mathbb{WCP}^1_{[n_1,n_2]}$.
Using the freedom to choose the flat connection in the KK reduction to $D=4$, we now set
\begin{align}
a^I =-\frac{\mathfrak{p}^I}{\mathfrak{p}^0} \ \Rightarrow \ \check{\mathfrak{p}}^I = 0 \ \Rightarrow \ \check{\Phi}^I_0|_0 = \check{\Phi}^I_0|_1\, .
\end{align}

On reducing the $D=5$ Killing spinor to $D=4$, the signs 
$\sigma_a$ defined via \eqref{KScharge} precisely become 
the signs defined in reference \cite{BenettiGenolini:2024hyd}, and which appear in equation~\eqref{Phiconstraint}. The chirality of the $D=4$ Killing spinor 
at the fixed point $x_a$ is then $\chi_a$, where $\chi_0=-\ca_0\ca_1$, 
$\chi_1=-\ca_1\ca_2$, $\chi_*=-\ca_2\ca_0$ \cite{BenettiGenolini:2024hyd}.

The constraint \eqref{Phiconstraint} is the same at all three fixed points $x_0,x_1,x_*$, and reads 
\begin{align}\label{4dconstraint}
	\ca_1 + \ca_2 \varepsilon_1+ \ca_0 \varepsilon_2 = - \zeta_I \check{\Phi}^I_0|_{0,1,*} \, .
\end{align}
Indeed, necessarily
\begin{align}
\check{\Phi}^I_0|_* = \check{\Phi}^I_0|_{0,1} \equiv \check{\Phi}^I\, ,
\end{align}
also holds.
One can see this by appealing to the fact that via a UV/IR 
equation  \cite{BenettiGenolini:2025icr,BenettiGenolini:2026qdm} the values $\check{\Phi}^I_0|_{0,1}$ on the horizon are related 
to quantities at the UV boundary, and that the same argument relates
$\check{\Phi}^I_0|_*$ to the \emph{same} boundary quantities. Alternatively,
using the compact glued space $\overline{M}_4=\mathbb{WCP}^2_{[p,n_1,n_2]}$ we have $H_2(\mathbb{WCP}^2_{[p,n_1,n_2]},\R)=\R$ has only one independent two-cycle, 
and since $\check{\mathfrak{p}}^I=0$ for the cycle at $z_*=0$ in \eqref{flux}, the same is true for the other
two spindle cycles at $z_0=0$, $z_1=0$, which connect the poles of the horizon to the point $x_*$.

Substituting the above into \eqref{I4FP_multi}, we find that all dependence 
on $\ell$ in \eqref{ell} indeed cancels out, giving  
\begin{align}\label{FPresults_v0}
	& I^{\rm FP}_{4\partial}  = \frac{\ii \pi^2}{2 G_{(5)}}\Bigg\{ - (C_{IJK} + \alpha  \tilde{\lambda}_{IJK}) \frac{  \check{\Phi}^I\check{\Phi}^J\check{\Phi}^K}{6\varepsilon_1 \varepsilon_2} \nn \\
	& \qquad \qquad + 4\alpha  \lambda_I \check{\Phi}^I  \Bigg[ (16\ttw - \frac{1}{2}\ttt) \frac{(\ca_1 + \ca_2 \varepsilon_1 + \ca_0 \varepsilon_2)^2}{\varepsilon_1 \varepsilon_2}\nonumber\\
& \qquad \qquad \qquad \qquad \qquad - 32\ttw \left( \frac{\varepsilon_1^2 + \varepsilon_2^2 + 1}{\varepsilon_1 \varepsilon_2} \right) \Bigg]\Bigg\} \,.
\end{align}
Using the constraint \eqref{4dconstraint}, the second line may be combined with the first line. Setting 
$\ttw=\frac{1}{32}$, $\ttt=\frac{1}{3}$, the values obtained from dimensional reduction of the $D=5$ action \eqref{eq:effective 4deriv lagrangian}, 
remarkably we find this precisely combines into 
$C^{(\alpha)}_{IJK}$ given in \eqref{CS_corrected}. The upshot is
\begin{align}
	I^{\rm FP}_{4\partial} & =  - \frac{\ii \pi^2}{12G_{(5)}}  {C_{IJK}^{(\alpha)}} \frac{\check\Phi^I \check\Phi^J \check\Phi^K}{\varepsilon_1 \varepsilon_2} \nonumber\\
& \qquad \qquad \qquad - \frac{2 \ii \pi^2}{G_{(5)}}  \alpha \lambda_I \check\Phi^I \left( \frac{\varepsilon_1^2 + \varepsilon_2^2 + 1}{\varepsilon_1 \varepsilon_2} \right) \,.
\end{align} 
This result and the constraint \eqref{4dconstraint} can be precisely matched with
the Cardy-like limit of the superconformal index \eqref{index} and \eqref{multi_chempot_constraint_v1}, respectively.
To do so we use the change of variables \eqref{AdS/CFT}, giving the map 
between 't Hooft anomaly coefficients and gravity variables, together with 
\begin{align}\label{mapy}
\omega_i \leftrightarrow 2\pi \ii \varepsilon_i\,, \quad \varphi^I \leftrightarrow 
-2\pi \ii  \check{\Phi}^I\, , \quad r_I\leftrightarrow  \frac{1}{2}\zeta_I\, .
\end{align}
To match, we have taken $\sigma_0=\sigma_2=+1$ which, from above, corresponds to the $D=4$ Killing
spinor having the same chirality at the poles of $\mathbb{WCP}^1_{[n_1,n_2]}$ (i.e. a topological twist), with $-\sigma_1$ 
 specifying minus the charge of the $D = 5$ Killing spinor around the thermal circle,
 matching the $\pm$ sign on the  right hand side of \eqref{multi_chempot_constraint_v1} \footnote{This is consistent with the explicit Killing spinor 
given in \cite{Cabo-Bizet:2018ehj},
which is charged under the Hopf (and uncharged under the anti-Hopf) direction in the $S^3$.
We can also take
$\sigma_0=\sigma_2=-1$, $\sigma_1$ is the $\pm$ sign on the right hand side of \eqref{multi_chempot_constraint_v1} and
with $r_I\leftrightarrow  -\frac{1}{2}\zeta_I$, associated with the conjugate Killing spinor.}.

\section{Discussion}
We have obtained a universal derivation of the Cardy-like limit of the superconformal index using localization
in supergravity. Our procedure uses a background subtraction prescription that attached a Euclidean
$AdS_5$ factor at the boundary which effectively leads to implementing
localization on a compact manifold without boundary. Interestingly, the final answer gets a contribution from
three fixed points after dimensional reduction, two arising from the black hole horizon and another from the centre of the 
$AdS_5$ subtraction factor \footnote{In the two-derivative computation, for the special choice of vector  
$\ell = (0,1,1)$, there is no contribution from the $AdS_5$ factor, but in the four-derivative computation there still is.
This possibility was not taken into account in \cite{Hu:2025ogz}.  }.
It would be desirable to obtain our result using holographic renormalization rather than background subtraction, but
there is still much to be understood regarding $D=5$ holographic renormalization before this can be attempted.

For the black hole solutions we were able to assume that the $D=5$ gauge fields are globally defined one-forms.
In order to generalize our results to compute higher derivative corrections to the on-shell actions for more general supersymmetric solutions,
including black rings and lenses (for related discussions at two-derivative level, see \cite{Cassani:2025iix,Boruch:2025sie,Colombo:2025yqy})
we need to relax this assumption \footnote{Alternatively, \cite{Colombo:2025ihp, Colombo:2025yqy} compute the two-derivative $D=5$ on-shell action using a 
``transverse'' form of the Berline--Vergne--Atiyah--Bott formula \cite{Goertsches:2015}.
We expect this approach to extend to higher derivatives, leading to 
equivalent formulas to those presented here.}. 
Finally, in this letter we have focused
on gauged supergravity, but our key results also apply to ungauged supergravity with additional applications. We aim to report on
these topics soon.

\section*{Acknowledgments}
We thank Davide Cassani and Pietro Benetti Genolini for discussions. This work was supported in part by STFC grants ST/X000575/1 and
ST/X000761/1.
JP is supported by a Dean's PhD studentship at Imperial College.
FG is supported by an STFC studentship.


\begin{thebibliography}{48}%
\makeatletter
\providecommand \@ifxundefined [1]{%
 \@ifx{#1\undefined}
}%
\providecommand \@ifnum [1]{%
 \ifnum #1\expandafter \@firstoftwo
 \else \expandafter \@secondoftwo
 \fi
}%
\providecommand \@ifx [1]{%
 \ifx #1\expandafter \@firstoftwo
 \else \expandafter \@secondoftwo
 \fi
}%
\providecommand \natexlab [1]{#1}%
\providecommand \enquote  [1]{``#1''}%
\providecommand \bibnamefont  [1]{#1}%
\providecommand \bibfnamefont [1]{#1}%
\providecommand \citenamefont [1]{#1}%
\providecommand \href@noop [0]{\@secondoftwo}%
\providecommand \href [0]{\begingroup \@sanitize@url \@href}%
\providecommand \@href[1]{\@@startlink{#1}\@@href}%
\providecommand \@@href[1]{\endgroup#1\@@endlink}%
\providecommand \@sanitize@url [0]{\catcode `\\12\catcode `\$12\catcode
  `\&12\catcode `\#12\catcode `\^12\catcode `\_12\catcode `\%12\relax}%
\providecommand \@@startlink[1]{}%
\providecommand \@@endlink[0]{}%
\providecommand \url  [0]{\begingroup\@sanitize@url \@url }%
\providecommand \@url [1]{\endgroup\@href {#1}{\urlprefix }}%
\providecommand \urlprefix  [0]{URL }%
\providecommand \Eprint [0]{\href }%
\providecommand \doibase [0]{https://doi.org/}%
\providecommand \selectlanguage [0]{\@gobble}%
\providecommand \bibinfo  [0]{\@secondoftwo}%
\providecommand \bibfield  [0]{\@secondoftwo}%
\providecommand \translation [1]{[#1]}%
\providecommand \BibitemOpen [0]{}%
\providecommand \bibitemStop [0]{}%
\providecommand \bibitemNoStop [0]{.\EOS\space}%
\providecommand \EOS [0]{\spacefactor3000\relax}%
\providecommand \BibitemShut  [1]{\csname bibitem#1\endcsname}%
\let\auto@bib@innerbib\@empty
\bibitem [{\citenamefont {Choi}\ \emph {et~al.}(2018)\citenamefont {Choi},
  \citenamefont {Kim}, \citenamefont {Kim},\ and\ \citenamefont
  {Nahmgoong}}]{Choi:2018hmj}%
  \BibitemOpen
  \bibfield  {author} {\bibinfo {author} {\bibfnamefont {S.}~\bibnamefont
  {Choi}}, \bibinfo {author} {\bibfnamefont {J.}~\bibnamefont {Kim}}, \bibinfo
  {author} {\bibfnamefont {S.}~\bibnamefont {Kim}},\ and\ \bibinfo {author}
  {\bibfnamefont {J.}~\bibnamefont {Nahmgoong}},\ }\bibfield  {title} {\bibinfo
  {title} {{Large AdS black holes from QFT}},\ }\href@noop {} {\  (\bibinfo
  {year} {2018})},\ \Eprint {https://arxiv.org/abs/1810.12067}
  {arXiv:1810.12067 [hep-th]} \BibitemShut {NoStop}%
\bibitem [{\citenamefont {Kim}\ \emph {et~al.}(2021)\citenamefont {Kim},
  \citenamefont {Kim},\ and\ \citenamefont {Song}}]{Kim:2019yrz}%
  \BibitemOpen
  \bibfield  {author} {\bibinfo {author} {\bibfnamefont {J.}~\bibnamefont
  {Kim}}, \bibinfo {author} {\bibfnamefont {S.}~\bibnamefont {Kim}},\ and\
  \bibinfo {author} {\bibfnamefont {J.}~\bibnamefont {Song}},\ }\bibfield
  {title} {\bibinfo {title} {{A 4d $ \mathcal{N} $ = 1 Cardy Formula}},\ }\href
  {https://doi.org/10.1007/JHEP01(2021)025} {\bibfield  {journal} {\bibinfo
  {journal} {JHEP}\ }\textbf {\bibinfo {volume} {01}},\ \bibinfo {pages}
  {025}},\ \Eprint {https://arxiv.org/abs/1904.03455} {arXiv:1904.03455
  [hep-th]} \BibitemShut {NoStop}%
\bibitem [{\citenamefont {Cabo-Bizet}\ \emph
  {et~al.}(2019{\natexlab{a}})\citenamefont {Cabo-Bizet}, \citenamefont
  {Cassani}, \citenamefont {Martelli},\ and\ \citenamefont
  {Murthy}}]{Cabo-Bizet:2019osg}%
  \BibitemOpen
  \bibfield  {author} {\bibinfo {author} {\bibfnamefont {A.}~\bibnamefont
  {Cabo-Bizet}}, \bibinfo {author} {\bibfnamefont {D.}~\bibnamefont {Cassani}},
  \bibinfo {author} {\bibfnamefont {D.}~\bibnamefont {Martelli}},\ and\
  \bibinfo {author} {\bibfnamefont {S.}~\bibnamefont {Murthy}},\ }\bibfield
  {title} {\bibinfo {title} {{The asymptotic growth of states of the 4d $
  \mathcal{N}=1 $ superconformal index}},\ }\href
  {https://doi.org/10.1007/JHEP08(2019)120} {\bibfield  {journal} {\bibinfo
  {journal} {JHEP}\ }\textbf {\bibinfo {volume} {08}},\ \bibinfo {pages}
  {120}},\ \Eprint {https://arxiv.org/abs/1904.05865} {arXiv:1904.05865
  [hep-th]} \BibitemShut {NoStop}%
\bibitem [{\citenamefont {Cassani}\ and\ \citenamefont
  {Komargodski}(2021)}]{Cassani:2021fyv}%
  \BibitemOpen
  \bibfield  {author} {\bibinfo {author} {\bibfnamefont {D.}~\bibnamefont
  {Cassani}}\ and\ \bibinfo {author} {\bibfnamefont {Z.}~\bibnamefont
  {Komargodski}},\ }\bibfield  {title} {\bibinfo {title} {{EFT and the SUSY
  Index on the 2nd Sheet}},\ }\href
  {https://doi.org/10.21468/SciPostPhys.11.1.004} {\bibfield  {journal}
  {\bibinfo  {journal} {SciPost Phys.}\ }\textbf {\bibinfo {volume} {11}},\
  \bibinfo {pages} {004} (\bibinfo {year} {2021})},\ \Eprint
  {https://arxiv.org/abs/2104.01464} {arXiv:2104.01464 [hep-th]} \BibitemShut
  {NoStop}%
\bibitem [{\citenamefont {Arabi~Ardehali}\ and\ \citenamefont
  {Murthy}(2021)}]{ArabiArdehali:2021nsx}%
  \BibitemOpen
  \bibfield  {author} {\bibinfo {author} {\bibfnamefont {A.}~\bibnamefont
  {Arabi~Ardehali}}\ and\ \bibinfo {author} {\bibfnamefont {S.}~\bibnamefont
  {Murthy}},\ }\bibfield  {title} {\bibinfo {title} {{The 4d superconformal
  index near roots of unity and 3d Chern-Simons theory}},\ }\href
  {https://doi.org/10.1007/JHEP10(2021)207} {\bibfield  {journal} {\bibinfo
  {journal} {JHEP}\ }\textbf {\bibinfo {volume} {10}},\ \bibinfo {pages}
  {207}},\ \Eprint {https://arxiv.org/abs/2104.02051} {arXiv:2104.02051
  [hep-th]} \BibitemShut {NoStop}%
\bibitem [{\citenamefont {Ohmori}\ and\ \citenamefont
  {Tizzano}(2022)}]{Ohmori:2021dzb}%
  \BibitemOpen
  \bibfield  {author} {\bibinfo {author} {\bibfnamefont {K.}~\bibnamefont
  {Ohmori}}\ and\ \bibinfo {author} {\bibfnamefont {L.}~\bibnamefont
  {Tizzano}},\ }\bibfield  {title} {\bibinfo {title} {{Anomaly matching across
  dimensions and supersymmetric Cardy formulae}},\ }\href
  {https://doi.org/10.1007/JHEP12(2022)027} {\bibfield  {journal} {\bibinfo
  {journal} {JHEP}\ }\textbf {\bibinfo {volume} {12}},\ \bibinfo {pages}
  {027}},\ \Eprint {https://arxiv.org/abs/2112.13445} {arXiv:2112.13445
  [hep-th]} \BibitemShut {NoStop}%
\bibitem [{\citenamefont {Cassani}\ \emph {et~al.}(2024)\citenamefont
  {Cassani}, \citenamefont {Ruip{\'e}rez},\ and\ \citenamefont
  {Turetta}}]{Cassani:2024tvk}%
  \BibitemOpen
  \bibfield  {author} {\bibinfo {author} {\bibfnamefont {D.}~\bibnamefont
  {Cassani}}, \bibinfo {author} {\bibfnamefont {A.}~\bibnamefont
  {Ruip{\'e}rez}},\ and\ \bibinfo {author} {\bibfnamefont {E.}~\bibnamefont
  {Turetta}},\ }\bibfield  {title} {\bibinfo {title} {{Higher-derivative
  corrections to flavoured BPS black hole thermodynamics and holography}},\
  }\href {https://doi.org/10.1007/JHEP05(2024)276} {\bibfield  {journal}
  {\bibinfo  {journal} {JHEP}\ }\textbf {\bibinfo {volume} {05}},\ \bibinfo
  {pages} {276}},\ \Eprint {https://arxiv.org/abs/2403.02410} {arXiv:2403.02410
  [hep-th]} \BibitemShut {NoStop}%
\bibitem [{Note1()}]{Note1}%
  \BibitemOpen
  \bibinfo {note} {Here $|\protect \mathcal {G}|$ is the order of a discrete
  one-form symmetry group, if such a symmetry exists, which would lead to a
  multiplicity of the saddle.}\BibitemShut {Stop}%
\bibitem [{\citenamefont {Cabo-Bizet}\ \emph
  {et~al.}(2019{\natexlab{b}})\citenamefont {Cabo-Bizet}, \citenamefont
  {Cassani}, \citenamefont {Martelli},\ and\ \citenamefont
  {Murthy}}]{Cabo-Bizet:2018ehj}%
  \BibitemOpen
  \bibfield  {author} {\bibinfo {author} {\bibfnamefont {A.}~\bibnamefont
  {Cabo-Bizet}}, \bibinfo {author} {\bibfnamefont {D.}~\bibnamefont {Cassani}},
  \bibinfo {author} {\bibfnamefont {D.}~\bibnamefont {Martelli}},\ and\
  \bibinfo {author} {\bibfnamefont {S.}~\bibnamefont {Murthy}},\ }\bibfield
  {title} {\bibinfo {title} {{Microscopic origin of the Bekenstein-Hawking
  entropy of supersymmetric AdS$_{5}$ black holes}},\ }\href
  {https://doi.org/10.1007/JHEP10(2019)062} {\bibfield  {journal} {\bibinfo
  {journal} {JHEP}\ }\textbf {\bibinfo {volume} {10}},\ \bibinfo {pages}
  {062}},\ \Eprint {https://arxiv.org/abs/1810.11442} {arXiv:1810.11442
  [hep-th]} \BibitemShut {NoStop}%
\bibitem [{\citenamefont {Cassani}\ and\ \citenamefont
  {Papini}(2019)}]{Cassani:2019mms}%
  \BibitemOpen
  \bibfield  {author} {\bibinfo {author} {\bibfnamefont {D.}~\bibnamefont
  {Cassani}}\ and\ \bibinfo {author} {\bibfnamefont {L.}~\bibnamefont
  {Papini}},\ }\bibfield  {title} {\bibinfo {title} {{The BPS limit of rotating
  AdS black hole thermodynamics}},\ }\href
  {https://doi.org/10.1007/JHEP09(2019)079} {\bibfield  {journal} {\bibinfo
  {journal} {JHEP}\ }\textbf {\bibinfo {volume} {09}},\ \bibinfo {pages}
  {079}},\ \Eprint {https://arxiv.org/abs/1906.10148} {arXiv:1906.10148
  [hep-th]} \BibitemShut {NoStop}%
\bibitem [{\citenamefont {Hosseini}\ \emph {et~al.}(2017)\citenamefont
  {Hosseini}, \citenamefont {Hristov},\ and\ \citenamefont
  {Zaffaroni}}]{Hosseini:2017mds}%
  \BibitemOpen
  \bibfield  {author} {\bibinfo {author} {\bibfnamefont {S.~M.}\ \bibnamefont
  {Hosseini}}, \bibinfo {author} {\bibfnamefont {K.}~\bibnamefont {Hristov}},\
  and\ \bibinfo {author} {\bibfnamefont {A.}~\bibnamefont {Zaffaroni}},\
  }\bibfield  {title} {\bibinfo {title} {{An extremization principle for the
  entropy of rotating BPS black holes in AdS$_{5}$}},\ }\href
  {https://doi.org/10.1007/JHEP07(2017)106} {\bibfield  {journal} {\bibinfo
  {journal} {JHEP}\ }\textbf {\bibinfo {volume} {07}},\ \bibinfo {pages}
  {106}},\ \Eprint {https://arxiv.org/abs/1705.05383} {arXiv:1705.05383
  [hep-th]} \BibitemShut {NoStop}%
\bibitem [{\citenamefont {Gutowski}\ and\ \citenamefont
  {Reall}(2004)}]{Gutowski:2004yv}%
  \BibitemOpen
  \bibfield  {author} {\bibinfo {author} {\bibfnamefont {J.~B.}\ \bibnamefont
  {Gutowski}}\ and\ \bibinfo {author} {\bibfnamefont {H.~S.}\ \bibnamefont
  {Reall}},\ }\bibfield  {title} {\bibinfo {title} {{General supersymmetric
  AdS(5) black holes}},\ }\href {https://doi.org/10.1088/1126-6708/2004/04/048}
  {\bibfield  {journal} {\bibinfo  {journal} {JHEP}\ }\textbf {\bibinfo
  {volume} {04}},\ \bibinfo {pages} {048}},\ \Eprint
  {https://arxiv.org/abs/hep-th/0401129} {arXiv:hep-th/0401129} \BibitemShut
  {NoStop}%
\bibitem [{\citenamefont {Cvetic}\ \emph {et~al.}(2004)\citenamefont {Cvetic},
  \citenamefont {Lu},\ and\ \citenamefont {Pope}}]{Cvetic:2004ny}%
  \BibitemOpen
  \bibfield  {author} {\bibinfo {author} {\bibfnamefont {M.}~\bibnamefont
  {Cvetic}}, \bibinfo {author} {\bibfnamefont {H.}~\bibnamefont {Lu}},\ and\
  \bibinfo {author} {\bibfnamefont {C.~N.}\ \bibnamefont {Pope}},\ }\bibfield
  {title} {\bibinfo {title} {{Charged rotating black holes in five dimensional
  $U(1)^3$ gauged N=2 supergravity}},\ }\href
  {https://doi.org/10.1103/PhysRevD.70.081502} {\bibfield  {journal} {\bibinfo
  {journal} {Phys. Rev. D}\ }\textbf {\bibinfo {volume} {70}},\ \bibinfo
  {pages} {081502} (\bibinfo {year} {2004})},\ \Eprint
  {https://arxiv.org/abs/hep-th/0407058} {arXiv:hep-th/0407058} \BibitemShut
  {NoStop}%
\bibitem [{\citenamefont {Witten}(1998)}]{Witten:1998qj}%
  \BibitemOpen
  \bibfield  {author} {\bibinfo {author} {\bibfnamefont {E.}~\bibnamefont
  {Witten}},\ }\bibfield  {title} {\bibinfo {title} {{Anti de Sitter space and
  holography}},\ }\href {https://doi.org/10.4310/ATMP.1998.v2.n2.a2} {\bibfield
   {journal} {\bibinfo  {journal} {Adv. Theor. Math. Phys.}\ }\textbf {\bibinfo
  {volume} {2}},\ \bibinfo {pages} {253} (\bibinfo {year} {1998})},\ \Eprint
  {https://arxiv.org/abs/hep-th/9802150} {arXiv:hep-th/9802150} \BibitemShut
  {NoStop}%
\bibitem [{\citenamefont {Bobev}\ \emph
  {et~al.}(2022{\natexlab{a}})\citenamefont {Bobev}, \citenamefont {Hristov},\
  and\ \citenamefont {Reys}}]{Bobev:2021qxx}%
  \BibitemOpen
  \bibfield  {author} {\bibinfo {author} {\bibfnamefont {N.}~\bibnamefont
  {Bobev}}, \bibinfo {author} {\bibfnamefont {K.}~\bibnamefont {Hristov}},\
  and\ \bibinfo {author} {\bibfnamefont {V.}~\bibnamefont {Reys}},\ }\bibfield
  {title} {\bibinfo {title} {{AdS$_{5}$ holography and higher-derivative
  supergravity}},\ }\href {https://doi.org/10.1007/JHEP04(2022)088} {\bibfield
  {journal} {\bibinfo  {journal} {JHEP}\ }\textbf {\bibinfo {volume} {04}},\
  \bibinfo {pages} {088}},\ \Eprint {https://arxiv.org/abs/2112.06961}
  {arXiv:2112.06961 [hep-th]} \BibitemShut {NoStop}%
\bibitem [{\citenamefont {Bobev}\ \emph
  {et~al.}(2022{\natexlab{b}})\citenamefont {Bobev}, \citenamefont {Dimitrov},
  \citenamefont {Reys},\ and\ \citenamefont {Vekemans}}]{Bobev:2022bjm}%
  \BibitemOpen
  \bibfield  {author} {\bibinfo {author} {\bibfnamefont {N.}~\bibnamefont
  {Bobev}}, \bibinfo {author} {\bibfnamefont {V.}~\bibnamefont {Dimitrov}},
  \bibinfo {author} {\bibfnamefont {V.}~\bibnamefont {Reys}},\ and\ \bibinfo
  {author} {\bibfnamefont {A.}~\bibnamefont {Vekemans}},\ }\bibfield  {title}
  {\bibinfo {title} {{Higher derivative corrections and AdS5 black holes}},\
  }\href {https://doi.org/10.1103/PhysRevD.106.L121903} {\bibfield  {journal}
  {\bibinfo  {journal} {Phys. Rev. D}\ }\textbf {\bibinfo {volume} {106}},\
  \bibinfo {pages} {L121903} (\bibinfo {year} {2022}{\natexlab{b}})},\ \Eprint
  {https://arxiv.org/abs/2207.10671} {arXiv:2207.10671 [hep-th]} \BibitemShut
  {NoStop}%
\bibitem [{\citenamefont {Cassani}\ \emph {et~al.}(2022)\citenamefont
  {Cassani}, \citenamefont {Ruip{\'e}rez},\ and\ \citenamefont
  {Turetta}}]{Cassani:2022lrk}%
  \BibitemOpen
  \bibfield  {author} {\bibinfo {author} {\bibfnamefont {D.}~\bibnamefont
  {Cassani}}, \bibinfo {author} {\bibfnamefont {A.}~\bibnamefont
  {Ruip{\'e}rez}},\ and\ \bibinfo {author} {\bibfnamefont {E.}~\bibnamefont
  {Turetta}},\ }\bibfield  {title} {\bibinfo {title} {{Corrections to AdS$_{5}$
  black hole thermodynamics from higher-derivative supergravity}},\ }\href
  {https://doi.org/10.1007/JHEP11(2022)059} {\bibfield  {journal} {\bibinfo
  {journal} {JHEP}\ }\textbf {\bibinfo {volume} {11}},\ \bibinfo {pages}
  {059}},\ \Eprint {https://arxiv.org/abs/2208.01007} {arXiv:2208.01007
  [hep-th]} \BibitemShut {NoStop}%
\bibitem [{Note2()}]{Note2}%
  \BibitemOpen
  \bibinfo {note} {$\alpha '$ corrections to the entropy for specific
  $AdS_5\times S^5$ black holes were shown to vanish directly in $D=10$ \cite
  {Melo:2020amq}.}\BibitemShut {Stop}%
\bibitem [{\citenamefont {Hanaki}\ \emph {et~al.}(2007)\citenamefont {Hanaki},
  \citenamefont {Ohashi},\ and\ \citenamefont {Tachikawa}}]{Hanaki:2006pj}%
  \BibitemOpen
  \bibfield  {author} {\bibinfo {author} {\bibfnamefont {K.}~\bibnamefont
  {Hanaki}}, \bibinfo {author} {\bibfnamefont {K.}~\bibnamefont {Ohashi}},\
  and\ \bibinfo {author} {\bibfnamefont {Y.}~\bibnamefont {Tachikawa}},\
  }\bibfield  {title} {\bibinfo {title} {{Supersymmetric Completion of an $R^2$
  term in Five-dimensional Supergravity}},\ }\href
  {https://doi.org/10.1143/PTP.117.533} {\bibfield  {journal} {\bibinfo
  {journal} {Prog. Theor. Phys.}\ }\textbf {\bibinfo {volume} {117}},\ \bibinfo
  {pages} {533} (\bibinfo {year} {2007})},\ \Eprint
  {https://arxiv.org/abs/hep-th/0611329} {arXiv:hep-th/0611329} \BibitemShut
  {NoStop}%
\bibitem [{\citenamefont {Benetti~Genolini}\ \emph {et~al.}(2023)\citenamefont
  {Benetti~Genolini}, \citenamefont {Gauntlett},\ and\ \citenamefont
  {Sparks}}]{BenettiGenolini:2023kxp}%
  \BibitemOpen
  \bibfield  {author} {\bibinfo {author} {\bibfnamefont {P.}~\bibnamefont
  {Benetti~Genolini}}, \bibinfo {author} {\bibfnamefont {J.~P.}\ \bibnamefont
  {Gauntlett}},\ and\ \bibinfo {author} {\bibfnamefont {J.}~\bibnamefont
  {Sparks}},\ }\bibfield  {title} {\bibinfo {title} {{Equivariant Localization
  in Supergravity}},\ }\href {https://doi.org/10.1103/PhysRevLett.131.121602}
  {\bibfield  {journal} {\bibinfo  {journal} {Phys. Rev. Lett.}\ }\textbf
  {\bibinfo {volume} {131}},\ \bibinfo {pages} {121602} (\bibinfo {year}
  {2023})},\ \Eprint {https://arxiv.org/abs/2306.03868} {arXiv:2306.03868
  [hep-th]} \BibitemShut {NoStop}%
\bibitem [{\citenamefont {Colombo}\ \emph
  {et~al.}(2025{\natexlab{a}})\citenamefont {Colombo}, \citenamefont
  {Dimitrov}, \citenamefont {Martelli},\ and\ \citenamefont
  {Zaffaroni}}]{Colombo:2025ihp}%
  \BibitemOpen
  \bibfield  {author} {\bibinfo {author} {\bibfnamefont {E.}~\bibnamefont
  {Colombo}}, \bibinfo {author} {\bibfnamefont {V.}~\bibnamefont {Dimitrov}},
  \bibinfo {author} {\bibfnamefont {D.}~\bibnamefont {Martelli}},\ and\
  \bibinfo {author} {\bibfnamefont {A.}~\bibnamefont {Zaffaroni}},\ }\bibfield
  {title} {\bibinfo {title} {{Equivariant localization in supergravity in odd
  dimensions}},\ }\href@noop {} {\  (\bibinfo {year} {2025}{\natexlab{a}})},\
  \Eprint {https://arxiv.org/abs/2502.15624} {arXiv:2502.15624 [hep-th]}
  \BibitemShut {NoStop}%
\bibitem [{\citenamefont {Colombo}\ \emph
  {et~al.}(2025{\natexlab{b}})\citenamefont {Colombo}, \citenamefont
  {Dimitrov}, \citenamefont {Martelli},\ and\ \citenamefont
  {Zaffaroni}}]{Colombo:2025yqy}%
  \BibitemOpen
  \bibfield  {author} {\bibinfo {author} {\bibfnamefont {E.}~\bibnamefont
  {Colombo}}, \bibinfo {author} {\bibfnamefont {V.}~\bibnamefont {Dimitrov}},
  \bibinfo {author} {\bibfnamefont {D.}~\bibnamefont {Martelli}},\ and\
  \bibinfo {author} {\bibfnamefont {A.}~\bibnamefont {Zaffaroni}},\ }\bibfield
  {title} {\bibinfo {title} {{Patch-wise localization with Chern-Simons forms
  in five dimensional supergravity}},\ }\href@noop {} {\  (\bibinfo {year}
  {2025}{\natexlab{b}})},\ \Eprint {https://arxiv.org/abs/2511.13824}
  {arXiv:2511.13824 [hep-th]} \BibitemShut {NoStop}%
\bibitem [{\citenamefont {Benetti~Genolini}\ \emph
  {et~al.}(2026{\natexlab{a}})\citenamefont {Benetti~Genolini}, \citenamefont
  {Gauntlett}, \citenamefont {Jiao}, \citenamefont {Park},\ and\ \citenamefont
  {Sparks}}]{BenettiGenolini:2025icr}%
  \BibitemOpen
  \bibfield  {author} {\bibinfo {author} {\bibfnamefont {P.}~\bibnamefont
  {Benetti~Genolini}}, \bibinfo {author} {\bibfnamefont {J.~P.}\ \bibnamefont
  {Gauntlett}}, \bibinfo {author} {\bibfnamefont {Y.}~\bibnamefont {Jiao}},
  \bibinfo {author} {\bibfnamefont {J.}~\bibnamefont {Park}},\ and\ \bibinfo
  {author} {\bibfnamefont {J.}~\bibnamefont {Sparks}},\ }\bibfield  {title}
  {\bibinfo {title} {{Equivariant localization for D = 5 gauged
  supergravity}},\ }\href {https://doi.org/10.1007/JHEP03(2026)080} {\bibfield
  {journal} {\bibinfo  {journal} {JHEP}\ }\textbf {\bibinfo {volume} {03}},\
  \bibinfo {pages} {080}},\ \Eprint {https://arxiv.org/abs/2508.08207}
  {arXiv:2508.08207 [hep-th]} \BibitemShut {NoStop}%
\bibitem [{\citenamefont {Benetti~Genolini}\ \emph {et~al.}(2024)\citenamefont
  {Benetti~Genolini}, \citenamefont {Gauntlett}, \citenamefont {Jiao},
  \citenamefont {L{\"u}scher},\ and\ \citenamefont
  {Sparks}}]{BenettiGenolini:2024xeo}%
  \BibitemOpen
  \bibfield  {author} {\bibinfo {author} {\bibfnamefont {P.}~\bibnamefont
  {Benetti~Genolini}}, \bibinfo {author} {\bibfnamefont {J.~P.}\ \bibnamefont
  {Gauntlett}}, \bibinfo {author} {\bibfnamefont {Y.}~\bibnamefont {Jiao}},
  \bibinfo {author} {\bibfnamefont {A.}~\bibnamefont {L{\"u}scher}},\ and\
  \bibinfo {author} {\bibfnamefont {J.}~\bibnamefont {Sparks}},\ }\bibfield
  {title} {\bibinfo {title} {{Localization of the Free Energy in
  Supergravity}},\ }\href {https://doi.org/10.1103/PhysRevLett.133.141601}
  {\bibfield  {journal} {\bibinfo  {journal} {Phys. Rev. Lett.}\ }\textbf
  {\bibinfo {volume} {133}},\ \bibinfo {pages} {141601} (\bibinfo {year}
  {2024})},\ \Eprint {https://arxiv.org/abs/2407.02554} {arXiv:2407.02554
  [hep-th]} \BibitemShut {NoStop}%
\bibitem [{\citenamefont {Benetti~Genolini}\ \emph
  {et~al.}(2025{\natexlab{a}})\citenamefont {Benetti~Genolini}, \citenamefont
  {Gauntlett}, \citenamefont {Jiao}, \citenamefont {L{\"u}scher},\ and\
  \citenamefont {Sparks}}]{BenettiGenolini:2024lbj}%
  \BibitemOpen
  \bibfield  {author} {\bibinfo {author} {\bibfnamefont {P.}~\bibnamefont
  {Benetti~Genolini}}, \bibinfo {author} {\bibfnamefont {J.~P.}\ \bibnamefont
  {Gauntlett}}, \bibinfo {author} {\bibfnamefont {Y.}~\bibnamefont {Jiao}},
  \bibinfo {author} {\bibfnamefont {A.}~\bibnamefont {L{\"u}scher}},\ and\
  \bibinfo {author} {\bibfnamefont {J.}~\bibnamefont {Sparks}},\ }\bibfield
  {title} {\bibinfo {title} {{Equivariant localization for D = 4 gauged
  supergravity}},\ }\href {https://doi.org/10.1007/JHEP08(2025)211} {\bibfield
  {journal} {\bibinfo  {journal} {JHEP}\ }\textbf {\bibinfo {volume} {08}},\
  \bibinfo {pages} {211}},\ \Eprint {https://arxiv.org/abs/2412.07828}
  {arXiv:2412.07828 [hep-th]} \BibitemShut {NoStop}%
\bibitem [{\citenamefont {Benetti~Genolini}\ \emph
  {et~al.}(2026{\natexlab{b}})\citenamefont {Benetti~Genolini}, \citenamefont
  {Gaar}, \citenamefont {Gauntlett},\ and\ \citenamefont
  {Sparks}}]{BenettiGenolini:2026qdm}%
  \BibitemOpen
  \bibfield  {author} {\bibinfo {author} {\bibfnamefont {P.}~\bibnamefont
  {Benetti~Genolini}}, \bibinfo {author} {\bibfnamefont {F.}~\bibnamefont
  {Gaar}}, \bibinfo {author} {\bibfnamefont {J.~P.}\ \bibnamefont
  {Gauntlett}},\ and\ \bibinfo {author} {\bibfnamefont {J.}~\bibnamefont
  {Sparks}},\ }\bibfield  {title} {\bibinfo {title} {{Equivariant localization
  for higher derivative supergravity}},\ }\href@noop {} {\  (\bibinfo {year}
  {2026}{\natexlab{b}})},\ \Eprint {https://arxiv.org/abs/2604.08656}
  {arXiv:2604.08656 [hep-th]} \BibitemShut {NoStop}%
\bibitem [{Note3()}]{Note3}%
  \BibitemOpen
  \bibinfo {note} {We also assume that any such solution, which in general is
  complex, is a \protect \emph {bona fide} saddle point of the Euclidean
  gravitational path integral for computing the superconformal index; for some
  further discussion see \cite
  {BenettiGenolini:2026raa,Krishna:2026rma}.}\BibitemShut {Stop}%
\bibitem [{\citenamefont {Ozkan}\ and\ \citenamefont
  {Pang}(2013)}]{Ozkan:2013nwa}%
  \BibitemOpen
  \bibfield  {author} {\bibinfo {author} {\bibfnamefont {M.}~\bibnamefont
  {Ozkan}}\ and\ \bibinfo {author} {\bibfnamefont {Y.}~\bibnamefont {Pang}},\
  }\bibfield  {title} {\bibinfo {title} {{All off-shell $R^{2}$ invariants in
  five dimensional $\mathcal{N} =$ 2 supergravity}},\ }\href
  {https://doi.org/10.1007/JHEP08(2013)042} {\bibfield  {journal} {\bibinfo
  {journal} {JHEP}\ }\textbf {\bibinfo {volume} {08}},\ \bibinfo {pages}
  {042}},\ \Eprint {https://arxiv.org/abs/1306.1540} {arXiv:1306.1540 [hep-th]}
  \BibitemShut {NoStop}%
\bibitem [{Note4()}]{Note4}%
  \BibitemOpen
  \bibinfo {note} {One can gauge fix in $D=5$ and then multiply by an overall
  factor of $\protect \frac {1}{16\pi G_{(5)}}$.}\BibitemShut {Stop}%
\bibitem [{\citenamefont {Hristov}\ \emph {et~al.}(2025)\citenamefont
  {Hristov}, \citenamefont {Khandelwal}, \citenamefont {Pang},\ and\
  \citenamefont {Tartaglino-Mazzucchelli}}]{Hristov:2025ygn}%
  \BibitemOpen
  \bibfield  {author} {\bibinfo {author} {\bibfnamefont {K.}~\bibnamefont
  {Hristov}}, \bibinfo {author} {\bibfnamefont {S.}~\bibnamefont {Khandelwal}},
  \bibinfo {author} {\bibfnamefont {Y.}~\bibnamefont {Pang}},\ and\ \bibinfo
  {author} {\bibfnamefont {G.}~\bibnamefont {Tartaglino-Mazzucchelli}},\
  }\bibfield  {title} {\bibinfo {title} {{Holographic origin of
  $a$-maximization and higher-derivative AdS$_5$/CFT$_4$}},\ }\href@noop {} {\
  (\bibinfo {year} {2025})},\ \Eprint {https://arxiv.org/abs/2511.22546}
  {arXiv:2511.22546 [hep-th]} \BibitemShut {NoStop}%
\bibitem [{\citenamefont {Butter}\ \emph {et~al.}(2014)\citenamefont {Butter},
  \citenamefont {de~Wit},\ and\ \citenamefont {Lodato}}]{Butter:2014iwa}%
  \BibitemOpen
  \bibfield  {author} {\bibinfo {author} {\bibfnamefont {D.}~\bibnamefont
  {Butter}}, \bibinfo {author} {\bibfnamefont {B.}~\bibnamefont {de~Wit}},\
  and\ \bibinfo {author} {\bibfnamefont {I.}~\bibnamefont {Lodato}},\
  }\bibfield  {title} {\bibinfo {title} {{Non-renormalization theorems and N=2
  supersymmetric backgrounds}},\ }\href
  {https://doi.org/10.1007/JHEP03(2014)131} {\bibfield  {journal} {\bibinfo
  {journal} {JHEP}\ }\textbf {\bibinfo {volume} {03}},\ \bibinfo {pages}
  {131}},\ \Eprint {https://arxiv.org/abs/1401.6591} {arXiv:1401.6591 [hep-th]}
  \BibitemShut {NoStop}%
\bibitem [{Note5()}]{Note5}%
  \BibitemOpen
  \bibinfo {note} {For the higher derivative theories, the boundary terms are
  under less control.}\BibitemShut {Stop}%
\bibitem [{Note6()}]{Note6}%
  \BibitemOpen
  \bibinfo {note} {Additional discussion of reality conditions in localization
  appear in \cite
  {BenettiGenolini:2024lbj,BenettiGenolini:2025icr,BenettiGenolini:2026qdm}.}\BibitemShut
  {Stop}%
\bibitem [{\citenamefont {Gibbons}\ and\ \citenamefont
  {Hawking}(1979)}]{Gibbons:1979xm}%
  \BibitemOpen
  \bibfield  {author} {\bibinfo {author} {\bibfnamefont {G.~W.}\ \bibnamefont
  {Gibbons}}\ and\ \bibinfo {author} {\bibfnamefont {S.~W.}\ \bibnamefont
  {Hawking}},\ }\bibfield  {title} {\bibinfo {title} {{Classification of
  Gravitational Instanton Symmetries}},\ }\href
  {https://doi.org/10.1007/BF01197189} {\bibfield  {journal} {\bibinfo
  {journal} {Commun. Math. Phys.}\ }\textbf {\bibinfo {volume} {66}},\ \bibinfo
  {pages} {291} (\bibinfo {year} {1979})}\BibitemShut {NoStop}%
\bibitem [{\citenamefont {Cassani}\ \emph {et~al.}(2025)\citenamefont
  {Cassani}, \citenamefont {Ruip{\'e}rez},\ and\ \citenamefont
  {Turetta}}]{Cassani:2025iix}%
  \BibitemOpen
  \bibfield  {author} {\bibinfo {author} {\bibfnamefont {D.}~\bibnamefont
  {Cassani}}, \bibinfo {author} {\bibfnamefont {A.}~\bibnamefont
  {Ruip{\'e}rez}},\ and\ \bibinfo {author} {\bibfnamefont {E.}~\bibnamefont
  {Turetta}},\ }\bibfield  {title} {\bibinfo {title} {{Bubbling saddles of the
  gravitational index}},\ }\href
  {https://doi.org/10.21468/SciPostPhys.19.5.134} {\bibfield  {journal}
  {\bibinfo  {journal} {SciPost Phys.}\ }\textbf {\bibinfo {volume} {19}},\
  \bibinfo {pages} {134} (\bibinfo {year} {2025})},\ \Eprint
  {https://arxiv.org/abs/2507.12650} {arXiv:2507.12650 [hep-th]} \BibitemShut
  {NoStop}%
\bibitem [{Note7()}]{Note7}%
  \BibitemOpen
  \bibinfo {note} {The gluing construction can be generalized to $\partial M_5
  = S^1_\tau \times M_3$, where $M_3$ are squashed three-spheres/lens spaces
  \cite {Park:2025fon}.}\BibitemShut {Stop}%
\bibitem [{Note8()}]{Note8}%
  \BibitemOpen
  \bibinfo {note} {The latter condition may be dropped, at the expense of
  introducing factors of $\protect \mathrm {gcd}(n_1,n_2)$ below.}\BibitemShut
  {Stop}%
\bibitem [{\citenamefont {Benetti~Genolini}\ \emph
  {et~al.}(2025{\natexlab{b}})\citenamefont {Benetti~Genolini}, \citenamefont
  {Gauntlett}, \citenamefont {Jiao}, \citenamefont {L{\"u}scher},\ and\
  \citenamefont {Sparks}}]{BenettiGenolini:2024hyd}%
  \BibitemOpen
  \bibfield  {author} {\bibinfo {author} {\bibfnamefont {P.}~\bibnamefont
  {Benetti~Genolini}}, \bibinfo {author} {\bibfnamefont {J.~P.}\ \bibnamefont
  {Gauntlett}}, \bibinfo {author} {\bibfnamefont {Y.}~\bibnamefont {Jiao}},
  \bibinfo {author} {\bibfnamefont {A.}~\bibnamefont {L{\"u}scher}},\ and\
  \bibinfo {author} {\bibfnamefont {J.}~\bibnamefont {Sparks}},\ }\bibfield
  {title} {\bibinfo {title} {{Toric gravitational instantons in gauged
  supergravity}},\ }\href {https://doi.org/10.1103/PhysRevD.111.046024}
  {\bibfield  {journal} {\bibinfo  {journal} {Phys. Rev. D}\ }\textbf {\bibinfo
  {volume} {111}},\ \bibinfo {pages} {046024} (\bibinfo {year}
  {2025}{\natexlab{b}})},\ \Eprint {https://arxiv.org/abs/2410.19036}
  {arXiv:2410.19036 [hep-th]} \BibitemShut {NoStop}%
\bibitem [{Note9()}]{Note9}%
  \BibitemOpen
  \bibinfo {note} {This is consistent with the explicit Killing spinor given in
  \cite {Cabo-Bizet:2018ehj}, which is charged under the Hopf (and uncharged
  under the anti-Hopf) direction in the $S^3$. We can also take $\sigma
  _0=\sigma _2=-1$, $\sigma _1$ is the $\pm $ sign on the right hand side of
  \protect \textup {\hbox {\mathsurround \z@ \protect \normalfont
  (\ignorespaces \ref {multi_chempot_constraint_v1}\unskip \@@italiccorr )}}
  and with $r_I\leftrightarrow -\protect \frac {1}{2}\zeta _I$, associated with
  the conjugate Killing spinor.}\BibitemShut {Stop}%
\bibitem [{Note10()}]{Note10}%
  \BibitemOpen
  \bibinfo {note} {In the two-derivative computation, for the special choice of
  vector $\ell = (0,1,1)$, there is no contribution from the $AdS_5$ factor,
  but in the four-derivative computation there still is. This possibility was
  not taken into account in \cite {Hu:2025ogz}.}\BibitemShut {Stop}%
\bibitem [{\citenamefont {Boruch}\ \emph {et~al.}(2025)\citenamefont {Boruch},
  \citenamefont {Emparan}, \citenamefont {Iliesiu},\ and\ \citenamefont
  {Murthy}}]{Boruch:2025sie}%
  \BibitemOpen
  \bibfield  {author} {\bibinfo {author} {\bibfnamefont {J.}~\bibnamefont
  {Boruch}}, \bibinfo {author} {\bibfnamefont {R.}~\bibnamefont {Emparan}},
  \bibinfo {author} {\bibfnamefont {L.~V.}\ \bibnamefont {Iliesiu}},\ and\
  \bibinfo {author} {\bibfnamefont {S.}~\bibnamefont {Murthy}},\ }\bibfield
  {title} {\bibinfo {title} {{Novel black saddles for 5d gravitational indices
  and the index enigma}},\ }\href@noop {} {\  (\bibinfo {year} {2025})},\
  \Eprint {https://arxiv.org/abs/2510.23699} {arXiv:2510.23699 [hep-th]}
  \BibitemShut {NoStop}%
\bibitem [{Note11()}]{Note11}%
  \BibitemOpen
  \bibinfo {note} {Alternatively, \cite {Colombo:2025ihp, Colombo:2025yqy}
  compute the two-derivative $D=5$ on-shell action using a ``transverse'' form
  of the Berline--Vergne--Atiyah--Bott formula \cite {Goertsches:2015}. We
  expect this approach to extend to higher derivatives, leading to equivalent
  formulas to those presented here.}\BibitemShut {Stop}%
\bibitem [{\citenamefont {Melo}\ and\ \citenamefont
  {Santos}(2021)}]{Melo:2020amq}%
  \BibitemOpen
  \bibfield  {author} {\bibinfo {author} {\bibfnamefont {J.~F.}\ \bibnamefont
  {Melo}}\ and\ \bibinfo {author} {\bibfnamefont {J.~E.}\ \bibnamefont
  {Santos}},\ }\bibfield  {title} {\bibinfo {title} {{Stringy corrections to
  the entropy of electrically charged supersymmetric black holes with
  $\mathrm{AdS}_5\times S^5$ asymptotics}},\ }\href
  {https://doi.org/10.1103/PhysRevD.103.066008} {\bibfield  {journal} {\bibinfo
   {journal} {Phys. Rev. D}\ }\textbf {\bibinfo {volume} {103}},\ \bibinfo
  {pages} {066008} (\bibinfo {year} {2021})},\ \Eprint
  {https://arxiv.org/abs/2007.06582} {arXiv:2007.06582 [hep-th]} \BibitemShut
  {NoStop}%
\bibitem [{\citenamefont {Benetti~Genolini}\ \emph
  {et~al.}(2026{\natexlab{c}})\citenamefont {Benetti~Genolini}, \citenamefont
  {Janssen},\ and\ \citenamefont {Murthy}}]{BenettiGenolini:2026raa}%
  \BibitemOpen
  \bibfield  {author} {\bibinfo {author} {\bibfnamefont {P.}~\bibnamefont
  {Benetti~Genolini}}, \bibinfo {author} {\bibfnamefont {O.}~\bibnamefont
  {Janssen}},\ and\ \bibinfo {author} {\bibfnamefont {S.}~\bibnamefont
  {Murthy}},\ }\bibfield  {title} {\bibinfo {title} {{Allowable complex metrics
  and the gravitational index of AdS$_5$ black holes}},\ }\href@noop {} {\
  (\bibinfo {year} {2026}{\natexlab{c}})},\ \Eprint
  {https://arxiv.org/abs/2601.23197} {arXiv:2601.23197 [hep-th]} \BibitemShut
  {NoStop}%
\bibitem [{\citenamefont {Krishna}\ and\ \citenamefont
  {Larsen}(2026)}]{Krishna:2026rma}%
  \BibitemOpen
  \bibfield  {author} {\bibinfo {author} {\bibfnamefont {V.}~\bibnamefont
  {Krishna}}\ and\ \bibinfo {author} {\bibfnamefont {F.}~\bibnamefont
  {Larsen}},\ }\bibfield  {title} {\bibinfo {title} {{Allowable Complex Black
  Holes in the Euclidean Gravitational Path Integral}},\ }\href@noop {} {\
  (\bibinfo {year} {2026})},\ \Eprint {https://arxiv.org/abs/2602.05979}
  {arXiv:2602.05979 [hep-th]} \BibitemShut {NoStop}%
\bibitem [{\citenamefont {Park}(2026)}]{Park:2025fon}%
  \BibitemOpen
  \bibfield  {author} {\bibinfo {author} {\bibfnamefont {J.}~\bibnamefont
  {Park}},\ }\bibfield  {title} {\bibinfo {title} {{Localizing AlAdS$_{5}$
  black holes and the SUSY index on S$^{1}${\texttimes} M$_{3}$}},\ }\href
  {https://doi.org/10.1007/JHEP06(2026)107} {\bibfield  {journal} {\bibinfo
  {journal} {JHEP}\ }\textbf {\bibinfo {volume} {06}},\ \bibinfo {pages}
  {107}},\ \Eprint {https://arxiv.org/abs/2511.15666} {arXiv:2511.15666
  [hep-th]} \BibitemShut {NoStop}%
\bibitem [{\citenamefont {Hu}\ \emph {et~al.}(2025)\citenamefont {Hu},
  \citenamefont {Hristov},\ and\ \citenamefont {Pang}}]{Hu:2025ogz}%
  \BibitemOpen
  \bibfield  {author} {\bibinfo {author} {\bibfnamefont {P.-J.}\ \bibnamefont
  {Hu}}, \bibinfo {author} {\bibfnamefont {K.}~\bibnamefont {Hristov}},\ and\
  \bibinfo {author} {\bibfnamefont {Y.}~\bibnamefont {Pang}},\ }\bibfield
  {title} {\bibinfo {title} {{Black hole thermodynamics at 4 derivatives,
  natural variables and BPS limits}},\ }\href
  {https://doi.org/10.1007/JHEP10(2025)118} {\bibfield  {journal} {\bibinfo
  {journal} {JHEP}\ }\textbf {\bibinfo {volume} {10}},\ \bibinfo {pages}
  {118}},\ \Eprint {https://arxiv.org/abs/2505.22726} {arXiv:2505.22726
  [hep-th]} \BibitemShut {NoStop}%
\bibitem [{\citenamefont {Goertsches}\ \emph {et~al.}(2017)\citenamefont
  {Goertsches}, \citenamefont {Nozawa},\ and\ \citenamefont
  {T{\"{o}}ben}}]{Goertsches:2015}%
  \BibitemOpen
  \bibfield  {author} {\bibinfo {author} {\bibfnamefont {O.}~\bibnamefont
  {Goertsches}}, \bibinfo {author} {\bibfnamefont {H.}~\bibnamefont {Nozawa}},\
  and\ \bibinfo {author} {\bibfnamefont {D.}~\bibnamefont {T{\"{o}}ben}},\
  }\bibfield  {title} {\bibinfo {title} {{Localization of Chern–Simons type
  invariants of Riemannian foliations}},\ }\href
  {https://doi.org/10.1007/s11856-017-1608-6} {\bibfield  {journal} {\bibinfo
  {journal} {Israel Journal of Mathematics}\ }\textbf {\bibinfo {volume}
  {222}},\ \bibinfo {pages} {867} (\bibinfo {year} {2017})},\ \Eprint
  {https://arxiv.org/abs/1508.07973} {arXiv:1508.07973 [math.DG]} \BibitemShut
  {NoStop}%
\end{thebibliography}

%

\end{document}